# Experimental verification of the feasibility of a quantum channel between Space and Earth

P. Villoresi[1], T. Jennewein[2], F. Tamburini[3], M. Aspelmeyer[2,4], C. Bonato[1],

R. Ursin[4], C. Pernechele[5], V. Luceri[6], G. Bianco[7], A. Zeilinger[2,4] and C. Barbieri[3]

[1]*Department of Information Engineering, University of Padova and*

*INFM-CNR LUXOR Laboratory for Ultraviolet and X-ray Optical Research, Padova, Italy*

[2]*Institute for Quantum Optics and Quantum Information (IQOQI), Austrian Academy of Sciences, Vienna, Austria*

[3]*Department of Astronomy, University of Padova, Italy*

[4]*Faculty of Physics, Institute for Experimental Physics, University of Vienna, Austria*

[5]*INAF-Cagliari, Capoterra (CA), Italy,*

[6]*Centro di Geodesia Spaziale "G. Colombo", e-GEOS S.p.A., Matera, Italy*

[7]*Centro di Geodesia Spaziale "G. Colombo", Agenzia Spaziale Italiana, Matera, Italy.*

*E-mail: paolo.villoresi@unipd.it*

**Abstract:** **Extending quantum communication to Space environments would enable to perform fundamental experiments on quantum physics as well as applications of quantum information at planetary and interplanetary scales. Here, we report the first experimental implementation of a scheme for single-photon exchange between a satellite and an Earth-based station. We built an experiment that mimics a single photon source on a satellite, exploiting the telescope at the Matera Laser Ranging Observatory of the Italian Space Agency to detect the transmitted photons. Weak laser pulses, emitted by the ground-based station, are directed towards a satellite equipped with cube-corner retroreflectors. These reflect a small portion of the pulse, with an average of less-than-one photon per pulse directed to our receiver, as required for the faint-pulse quantum communication. We were able to detect returns from satellite Ajisai, a low-earth orbit geodetic satellite, whose orbit has a perigee height of 1485 km.**

# 1. Introduction

Free-space quantum communication[1, 2], with fixed transmitter and receivers, has been successfully implemented, with links at increasing distances[3-9], reaching 144 Km in a recent experiment[10]. However the extension to even longer distances, in order to perform global-scale distribution of secret keys, or fundamental quantum mechanics experiments, is problematic due to the fact that the long-distance propagation of the optical beam is affected by several atmospheric and geographical hurdles. The most critical issues are wavefront distortion due to atmospheric turbulence, the stability of the optical systems against environmental vibrations and the quality of the large optics needed to mimize the diffractive spreading of the beam diameter. Further constraints follow from Earth's geography (curvature, orography, obstacles, etc.), and from the absorption of radiation in the lower layers of the atmosphere. Since the portion of atmosphere crossed by photons in a Zenith pass corresponds to an equivalent horizontal path of only 8 km at sea-level, we may note that orbiting satellites equipped with quantum terminals could overcome many of the above difficulties, establishing quantum communication links with ground-based stations on arbitrary locations around the Globe[11-18].

Here we report the experimental investigations on the exchange of single photons between a low Earth orbiting (LEO) transmitter based on a laser-ranged satellite and a ground-based receiver. These systems have been devised for geodynamical studies and are used to monitor the Earth's gravitational field by means of a series of measurements of the round-trip time (range) of an optical pulse from a station on Earth and the retroreflectors on the satellites[19]. This technique is known as Satellite Laser Ranging (SLR). We took advantage of the fact that their trajectories can be predicted in real time with good precision thanks to very accurate modeling based on their previous passes observed by the International Laser Ranging Service (ILRS)[20] network.

In our experiment we simulated a single photon source on a satellite using the retroreflection of a weak laser pulse from a SLR satellite: we chose the relevant experimental parameters in order to bring the number of photons per laser pulse in the downward link much less than unity. Our investigation differs with respect to the SLR techniques, even when the latter reaches the single-photon regime as in the case of Moon laser-ranging or kHz SLR[24], since we are counting the returns in a series of pre-determined time bins and not measuring the range time. Our observable is not the range itself but the number of detected photons per second, the detector count rate (DCR), as an initial step towards the measurement of the individual photons in quantum communication.

Our aim is to demonstrate that a source corresponding to a single photon emitter on a LEO satellite can indeed be identified and detected by an optical ground station, against a very high background noise, comparing the response for different satellites at different distances from the Earth surface.

## 2. Scheme of the experiment

The scheme of the ground setup is shown in Fig 1 and consists of a source equipped with a pulsed laser (wavelength 532 nm, repetition rate 17 kHz, energy per pulse 490 nJ and duration of 700 ps) installed along the Coudè path of the 1.5 m telescope of the Matera Laser Ranging Observatory (MLRO)[21] of the Italian Space Agency ASI, located in Matera, Italy. The pulse is directed toward the satellite (Uplink), as indicated in Fig. 2. The portion of the radiation retroreflected into the field-of-view (FOV) of the telescope (Downlink) constitutes the single-photon channel. The key point in the experiment is the use of weak pulses, whose energy is significantly lower than that of the MLRO (100 mJ) as well as the other known SLR sources[25]. In this way, by means of the link budget calculations we can assess that the down link is indeed in the single-photon regime, which mimics a satellite quantum communication source.

In our experiment an arbitrary polarization state of the photons cannot be preserved, due to multiple reflections along the optical path as well as the backreflection by the satellite[18]: only a dedicated source, with active polarization tracking and control would allow to use the polarization state of the transmitted photons for quantum communication purposes. However, the circular state of polarization is preserved in will be flipped to the orthogonal circular state under reflection. Therefore the circular polarization is used to separate the outgoing from the incoming light beam.

In order to properly track the satellite, it was first acquired and tracked by the MLRO original laser ranging system. The range values were measured using its strong laser pulses (with a repetition rate of 10 Hz, wavelength 532 nm, pulse energy 100 mJ). After several tens of seconds of successful SLR tracking, the telescope optical path was switched onto the quantum-channel system by moving a motorized mirror into the beam, while the telescope continued to follow in open loop the satellite along its trajectory. The quantum-channel system was kept active for an interval of a few minutes, and then the configuration was switched back to the MLRO system, in order to check the satellite tracking status and to perform further laser-ranging measurements. Typically, two or more alternations between the MLRO ranging system and our quantum channel were done during each pass of the satellite. Ref. 27 provides further details, including the optical design and detector characteristics of our quantum receiver/transmitter.

## 3. Estimating the satellite link budget

The expected detector count rate DCR for the different satellites were calculated according to Degnan radar link equation[23]

$$DCR = \Lambda\, \eta_d\, N_0\, \eta_t\, G_t\, \sigma \left( \frac{1}{4\pi R^2} \right)^2 A_t\, \eta_r\, T_a^2\, T_c^2 \ ,$$

where the various experimental parameters are defined in Table 1, based on the data for the Ajisai satellite obtained from the ILRS database[26].

We modeled the link losses and expected detector rates for the satellites under investigation, and the results are listed in Table 2. From this table it is evident that for all satellites we are in the quantum regime, in the sense that the number of photons in the channel per laser shot is much less than one. However there is a large difference in the expected count rates from the different satellites, making Ajisai the best choice for establishing the link according to our needs.

The geodetic satellites considered for our experiment were the most visible ones from MLRO: Ajisai (perigee height of 1485 km), Lageos II (5625 km), Topex-Poseidon (1350 km) and Beacon-C (927 km). The number and type of the retroreflectors as well as the orbits vary significantly among these satellites, leading to different expected optical responses, among which that of Ajisai has the most favourable one.

## 4. Adaptation of MLRO system for quantum communication

Since the very nature of our experiment is to analyze a train of single photons as compared to the standard timing analysis of usual SLR, we had to adapt the MLRO system for our requirements. To discriminate the photons belonging to the single photon link reflected by the satellite, from the large number of background photons entering the receiver, the incoming beam was filtered with respect to direction, wavelength and polarization of the incoming beam, and photons time of arrival.

For the directional filtering, the design of the FOV has to mediate between different requirements: on one hand it is desirable to narrow the field in order to reduce the amount of background, which scales with the subtended solid angle. On the other hand, an appropriately large FOV is necessary to account for the varying arriving angle of the single photons with respect to the outgoing beam due to velocity aberration[28], and to the blur due to atmospheric fluctuations and the pointing noise of the telescope. We found that the optimal FOV value for our experimental conditions was 30 arcseconds. In order to maximize the rejection of the spurious light, an interference band-pass filter with large angular acceptance was used. Moreover, a low-scattering Glan polarizer and a quarter wave plate were also used, in order to cross the polarizations of the up- and down-ward beams. This indeed results as the optimal choice for the rejection of the background photons and good contrast in the calibrations.

As regards the time of arrival, the round-trip duration was calculated for each laser shot, on the basis of the satellite distance (range, $R_s$) obtained using the predicted satellite position distributed by ILRS. All time-sensitive operations (laser firings, detector tags, telescope tracking increments, etc.) were referred by our FPGA-based timetagging unit to the UTC provided by the atomic clock at the MLRO, to which are synchronized the ILRS ephemerides. To further refine the *a-posteriori* analysis, the range values were improved by means of the NASA/GSFC analysis code Geodyn[22] which applies a very complete modeling of the forces acting on the satellite and uses all the observations by the worldwide ILRS network, including those acquired by MLRO operating in its nominal mode. A timing accuracy on the orbital data equal or better than 1 ns was thus achieved, ensuring that the estimate of the arrival time in the data analysis was derived by the best available information for the satellite pass.

# 5. Analysis of the detection events

All detection events were correlated with the transmitted laser pulses: for each detection event time stamp $t_{ret}$, the deviation D from the expected return time $t_{exp}$, D = $t_{exp}$ - $t_{ret}$ was then computed. These D values, grouped in several bins $\Delta t$ of varying width (from 1 to 20 ns), were accumulated over short arcs of the total satellite pass.

According to our criteria, the evidence of the single photon link would have been provided by a statistically significant peak (higher than 3 standard deviations $\sigma$ of the D value distribution) centred at D = 0. Furthermore, this peak had to persist in the histograms for various values of $\Delta t$, to prove that it was not a statistical artefact of the histogram.

A peak in the returns' histograms satisfying both conditions was identified for the Ajisai pass, see Fig 3, in the time interval 11-16 s after the start of acquisition. The peak is centered at D = 0, as required, and is nearly 5$\sigma$ above the mean. Moreover, the peak remains well above the statistical limit of 3 standard deviations $\sigma$ even when analyzing the data with various time bins $\Delta t$ between 3 to 19 ns, as shown in Fig. 4. Its statistical significance is at its highest value for $\Delta t$ = 5 ns. This value nicely confirms our expectation: the lower side is set by the addition of the instrumental jitters, the higher side by lowering the S/N ratio for accepting a higher spurious background at large bin sizes.

Therefore, the statistical significance of this peak is clearly established. The measured count rate in the peak is 5 counts per second, corresponding to a probability to detect a photon per emitted laser pulse of $3 \times 10^{-4}$. Taking into account the losses due to detection efficiency (-10dB) and the losses in the detection path (-11dB), the average photon number per pulse, $\mu$, emitted by the satellite and acquired by our detector is approximately $4 \times 10^{-2}$, i.e. well within the single photon regime.

## 6. Characterization of timing accuracy

The accurate determination of the arrival time for the photons proved to be the most crucial part in the analysis of the results. The timing jitter was characterized by measuring the range value for ground targets around MLRO, and found to be about 1 ns. A detailed analysis of the sources of the temporal jitter have shown that they can be ascribed mainly to the timing of the outgoing laser pulse (0.5 ns), the single-photon detection (0.6 ns) the temporal stamping (0.2 ns) and the internal clock stability (0.2 ns).

Additional sources of time jitter, each having approximately the same magnitude as the instrumental deficiencies, are the error in the synchronization with the timing signal of the station, and the precision in the orbital predictions. Regarding the influence of the satellite orbit on the arrival time: the instantaneous range $R_s$ is of the order of $R_s \approx 10^6$-$10^7$ m. Since the satellite is moving, the duration of the photon travel from the source to the receiver, which is proportional to $R_s$, varies very rapidly, up to several $\mu$s/second. Each satellite pass lasts from several minutes to half an hour, according to the satellite orbit, as shown for example in the case of Ajisai in Fig. 5. During this interval, the satellite trajectory is far from being monotonic, being influenced by the Earth's gravity anomalies, which introduce rapid variations of the range value as high as 100 ns. Therefore, the time delay between emission and arrival times is a rapidly changing and difficult to model quantity, but its proper determination is of crucial importance for the realization of a quantum channel. Accordingly, we expected that the best signal to noise ratio could be achieved for bin sizes $\Delta t$ from 1 to 20 ns, and this expectation was confirmed by the subsequent analysis.

# 7. Results and Conclusions

According to our statistical analysis, we have been able to detect returned photons, and to measure a return rate of about 5 counts per second for the Ajisai satellite. For each laser shot, about $1.2 \ 10^5$ photons leave the satellite in the whole solid angle subtended by the downlink field-of-view, indicated in Fig. 2. According to the link-budget equation, only an average of 0.4 photons are directed in the downward channel, thus realizing the condition of the single-photon channel. For Beacon C, Topex and Lageos II, the expected count-rate were about 1.1, 0.8 and 0.01 pps, and resulted too weak to significantly overcome the fluctuations of the background.

The return rate observed for Ajisai corresponds to a total attenuation along the light path of -157 dB, including the telescope, the optical bench and the detector.

We estimate for the return rate and for its fluctuation the value of 0.4 counts-per-second, cps, and the rms of 0.1 cps, in clear night conditions and within the 30 arcseconds of the gathered field-of-view (FOV), in a bandwidth of 1 nm and within the optimal time-bin width of 5 ns. The fluctuations of the background directly influence the possibility to detect the returned signal, and have been carefully minimized by shielding the receiver and gating off the detector the laser pulse reflections internal to the setup.

Our findings strongly underline the feasibility of the Space-to-Earth quantum communication with available technology. Indeed, by considering a spaceborne entangled source equipped with a 20 cm transmitting telescope and a 60 cm receiver, the one-way link losses are expected to be lower than 20 dB[13]. From our direct measure of the background, by considering a 3 ns time slot and a 0.15 nm spectral filtering, the fluctuation level is of 0.01 pps

A very important further achievement in our experiment is that we demonstrated the adaptation of a state-of-the-art satellite laser ranging observatory to act as a quantum communication transceiver. Obviously our experiment made use of the high tracking capability of this facility, but also extensively benefited from the existing infrastructure such as the real-time satellite prediction, the timing signals, the synchronized time-base, and also the International Laser Ranging Service including accurate orbital data of satellites.

This experiment represents a crucial step towards future space-to-ground quantum communication. Compared to other possible schemes with the source on the ground, this is clearly the preferred scenario[13,14].

In this work, we have achieved significant experimental results towards the realization of a quantum communication channel from a source on a LEO satellite to the receiver on Earth, as well as how to actually adapt a existing laser ranging facility for quantum communication. The high level of losses in the experiment in the two-way path prevented us from the implementation of a quantum-key-distribution protocol. Nevertheless, the good agreement of the return data with the model for the satellite link and the effective detection of returns by using present technology that result here also corroborated by the findings of the entanglement-based communication in the very long terrestrial link of 144 km[10], attest clearly on an experimental base the feasibility of satellite-based quantum channel in the near future. The configuration of such a satellite system has to keep the strict requirements of an emission of less than a single photon-per-shot, maintain the polarization state for the transmission during the satellite passage over the ground observer, and perform onboard timing and synchronization tasks. The optical losses and the timing of the emitted pulses are further issues have been studied in this work, and the approach towards effectively implementing the quantum link is clear. Finally, we have shown here that the tracking and photon-gathering capability of a ground station such as MLRO can be considered highly suitable for a space to ground quantum link.

## Acknowledgements

The authors kindly acknowledge the financial support by the University of Padova (Advanced Research Projects), by the Austrian Science Fund (FWF), the Austrian Research Promotion Agency (FFG) under ASAP-I and ASAP-II, and by the European Commission under the Integrated Project Qubit Applications (QAP) funded by the IST directorate. The authors also kindly acknowledge the International Laser Ranging Service for providing satellite orbital data, the help from the staff of MLRO and the advices of Dr. C. Romualdi of CRIBI, and Dr T. Occhipinti and Prof. G. Cariolaro and Prof A. Chiuso of DEI, University of Padova.

The authors declare no competing financial interests.

**Figures**

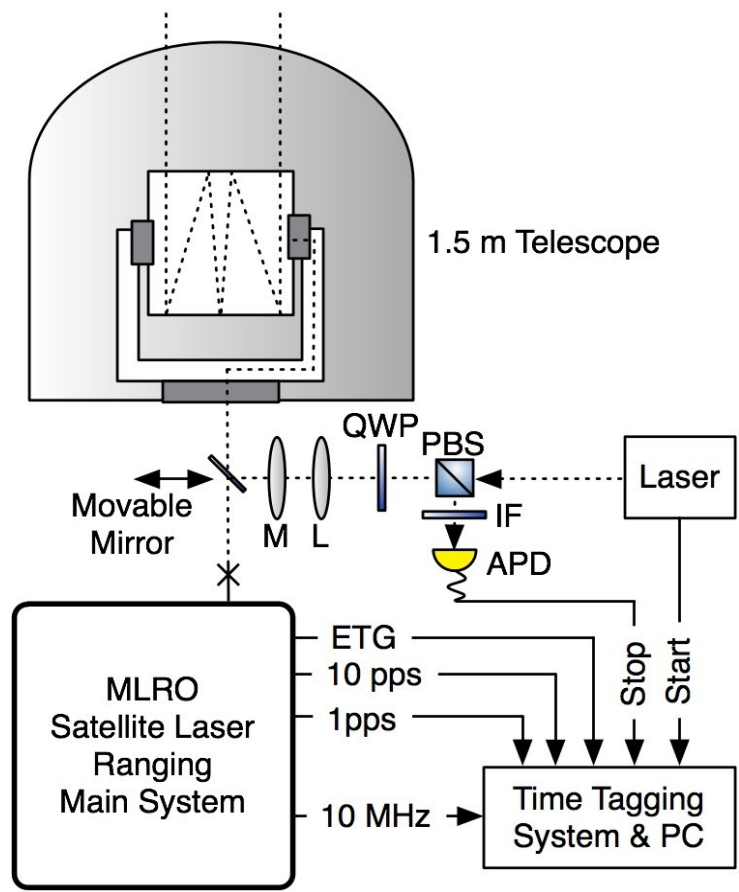

**Figure 1** - The experimental setup installed at the Matera Laser Ranging Observatory (MLRO): a train of pulses of 700-ps duration, 532 nm wavelength, 490 nJ of energy and 17 kHz repetition rate is sent toward the satellite via MLRO telescope, whose primary mirror has 1.5 m aperture. A fraction of the photons in the downlink path are collected by the telescope and detected by a gated silicon avalanche photo diode detector (D), placed behind a spectral filter (F). The separation in polarization of the uplink and downlink paths is accomplished by a polarizing beam-splitter (PBS) and a quarter wave-plate (QWP). All relevant events in the time domain, e.g. laser emission, clicks of the detector and timing signals ( 1pps, 10pps from MLRO atomic clock and the Event-Time-Gate ETG) are recorded as time-tags, and then referred to Coordinated Universal Time (UTC).

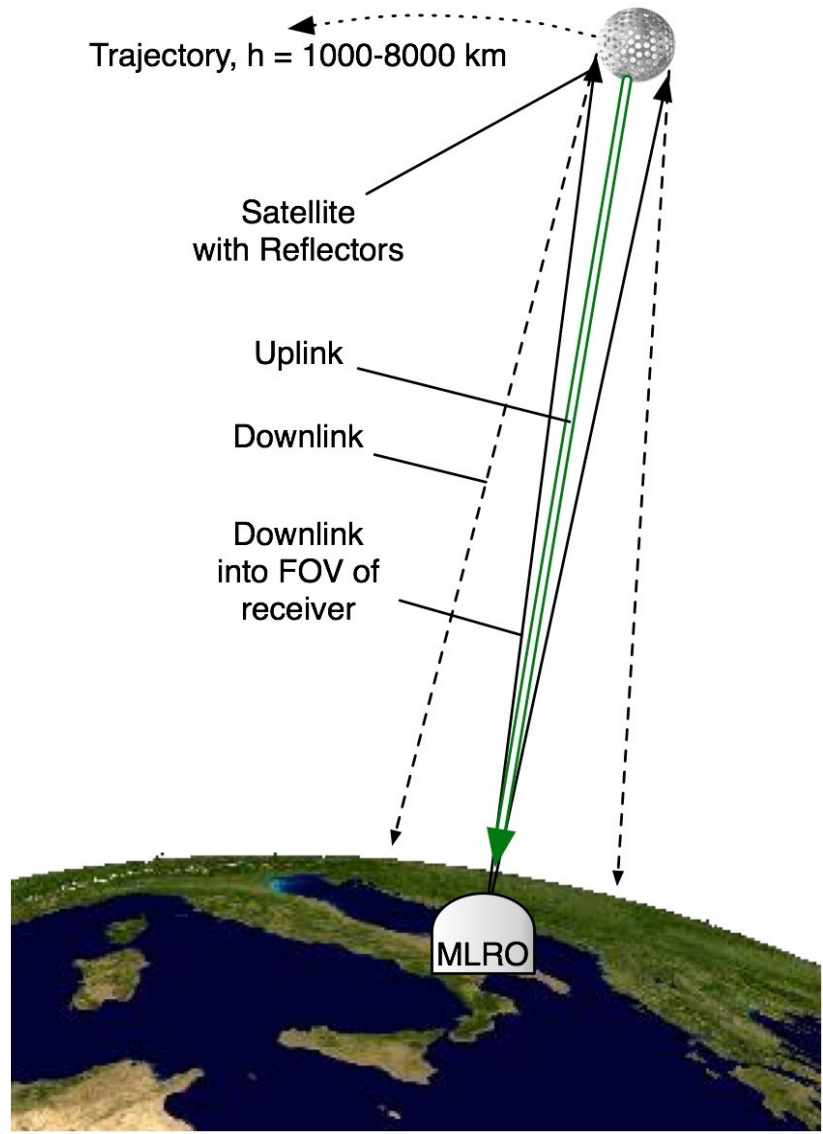

**Figure 2 -** The scheme of the satellite single-photon link. A fraction of the beam in the uplink path irradiates the satellite (LAGEOS shown). The corner cubes on the satellite retro-reflect back to Earth a small portion of the photons in the laser pulse (Downlink) and the gathered portion, according to the receiver FOV is indicated in green. This, according to the experimental parameters, is the single-photon channel.

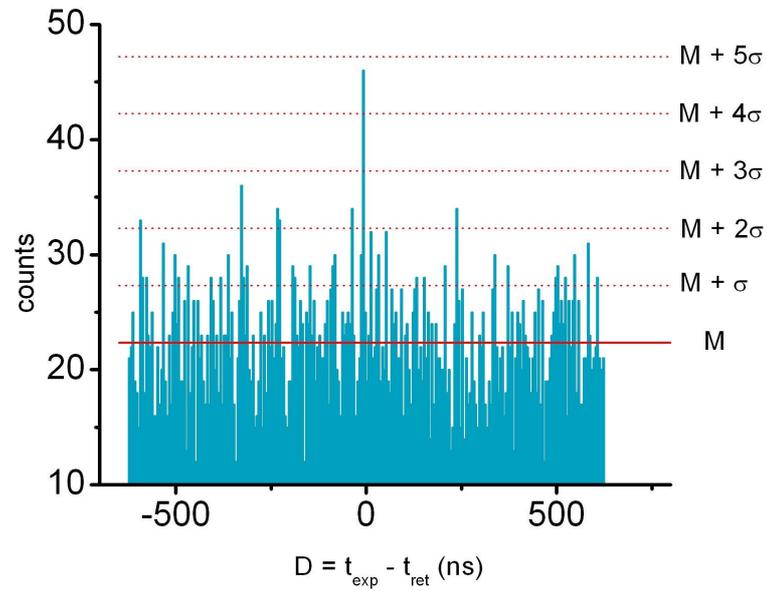

**Figure 3 -** Histogram of the differences D between expected and observed detections for Ajisai satellite. The peak of the histogram is centered at D = t$_{exp}$ - t$_{ret}$ = 0 ns, as expected, and is larger than the mean value of the background counts by 4.5 standard deviations. The bin size is $\Delta$t = 5 ns.

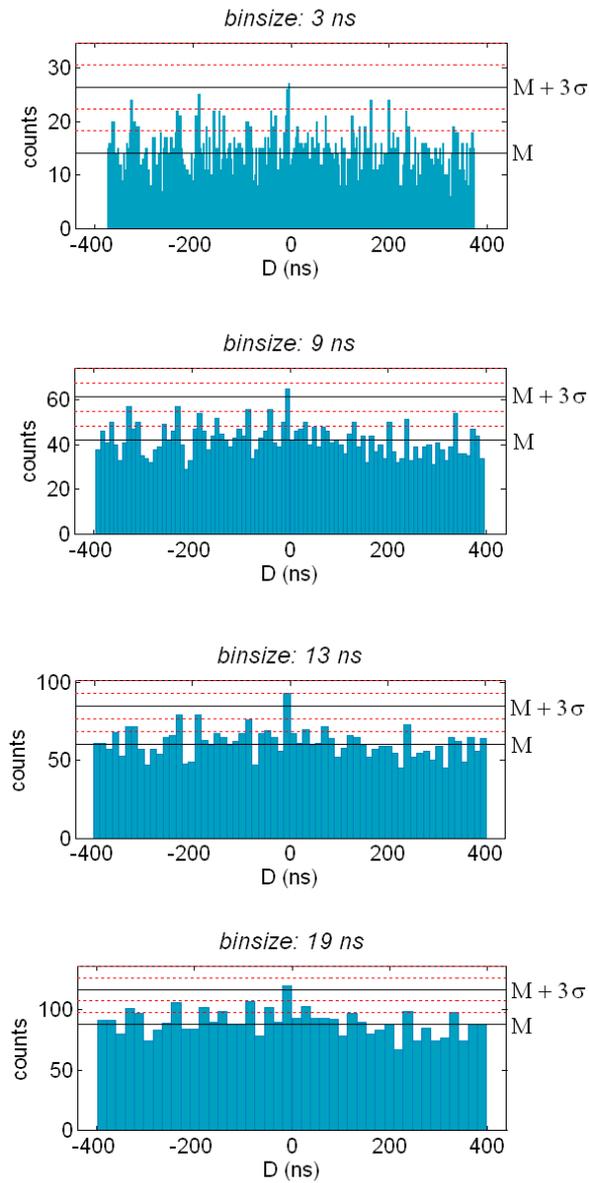

**Figure 4** – Consistency of the quantum channel identification: the central peak of the histogram of the detected events clearly persists above the noise level while the bin size Δt is changed from 3 to 19 ns.

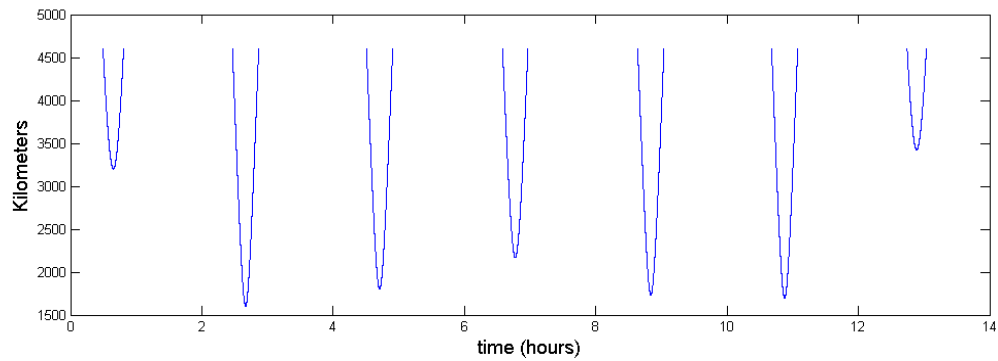

**Figure 5** - Plot of the range $R_s$ between the satellite and MLRO, for a typical series of passes of the AJISAI satellite. The round-trip time of the signal, directly proportional to $R_s$, exhibits strong and rapid variations during the satellite pass. This is accompanied by irregularities of different origins that require careful modeling in order to identify the 'good' photons in the quantum link.

## Tables

| | | |
|---|---|---|
| Laser repetition rate | $\Lambda$ | 17 KHz |
| Energy per laser pulse | $E_0$ | 490 nJ |
| Photons per emitted pulse | $N_0$ | $2.3 \cdot 10^{16}$ |
| Detector quantum efficiency | $\eta_d$ | 0.1 |
| Efficiency transmitting optics | $\eta_T$ | $5 \cdot 10^{-3}$ |
| Transmitter gain | $G_T$ | $7 \cdot 10^9$ |
| Satellite cross-section | $\sigma$ | $1.2 \cdot 10^7 \, m^2$ |
| Satellite slant distance | $R$ | 1650 km |
| Telescope area | $A_T$ | $1.77 \, m^2$ |
| Efficiency of receiving optics | $\eta_R$ | $4 \cdot 10^{-3}$ |
| One-way atmospheric transmission | $T_A$ | $8.1 \cdot 10^{-1}$ |
| One-way cirrus transmission | $T_C$ | 1 |
| **Expected detector rate** | | **4.6 cps** |

**Table 1** – Values of the experimental parameters for the Ajisai satellite during the acquisition.

| | Ajisai | Beacon | Topex | Lageos |
|---|---|---|---|---|
| detector rate | 4.6 | 1.2 | 0.8 | 0.01 |
| loss in downlink | $2.2 \cdot 10^{-9}$ | $6.6 \cdot 10^{-10}$ | $3.7 \cdot 10^{-10}$ | $8.4 \cdot 10^{-11}$ |
| photons in the channel per shot | 0.38 | 0.10 | 0.06 | 0.0009 |

**Tables 2** – Comparison of expected detector count probabilities for retroflected photons from the various satellites under investigation.